\documentclass[11pt,a4paper,superscriptaddress,prb,amsmath,amssymb,aps]{revtex4}

\usepackage{graphicx}
\usepackage{bm}
\usepackage[all]{xy}

\newcommand{\qed}{\nobreak \ifvmode \relax \else
      \ifdim\lastskip<1.5em \hskip-\lastskip
      \hskip1.5em plus0em minus0.5em \fi \nobreak
      \vrule height0.75em width0.5em depth0.25em\fi}

\begin{document}
\title{Robust quantization of a molecular motor motion in a stochastic environment.}

\author{V.~Y. Chernyak}
\affiliation{Department of Chemistry, Wayne State University, 5101 Cass Avenue, Detroit, MI 48202}
\affiliation{Theoretical Division, Los Alamos National Laboratory, Los Alamos, NM 87545 USA}
\author{N.~A. Sinitsyn}
\affiliation{Theoretical Division, Los Alamos National Laboratory, Los Alamos, NM 87545 USA}


\begin{abstract}
We explore quantization of the response of a molecular motor to periodic modulation of
control parameters. We formulate the Pumping-Quantization Theorem (PQT) that identifies the conditions for robust integer  quantized behavior
of a periodically driven molecular machine. Implication of PQT on experiments with catenane
molecules are discussed.
\end{abstract}

\date{\today}

\maketitle

\begin{figure}[t]
\centerline{\includegraphics[width=2.8in]{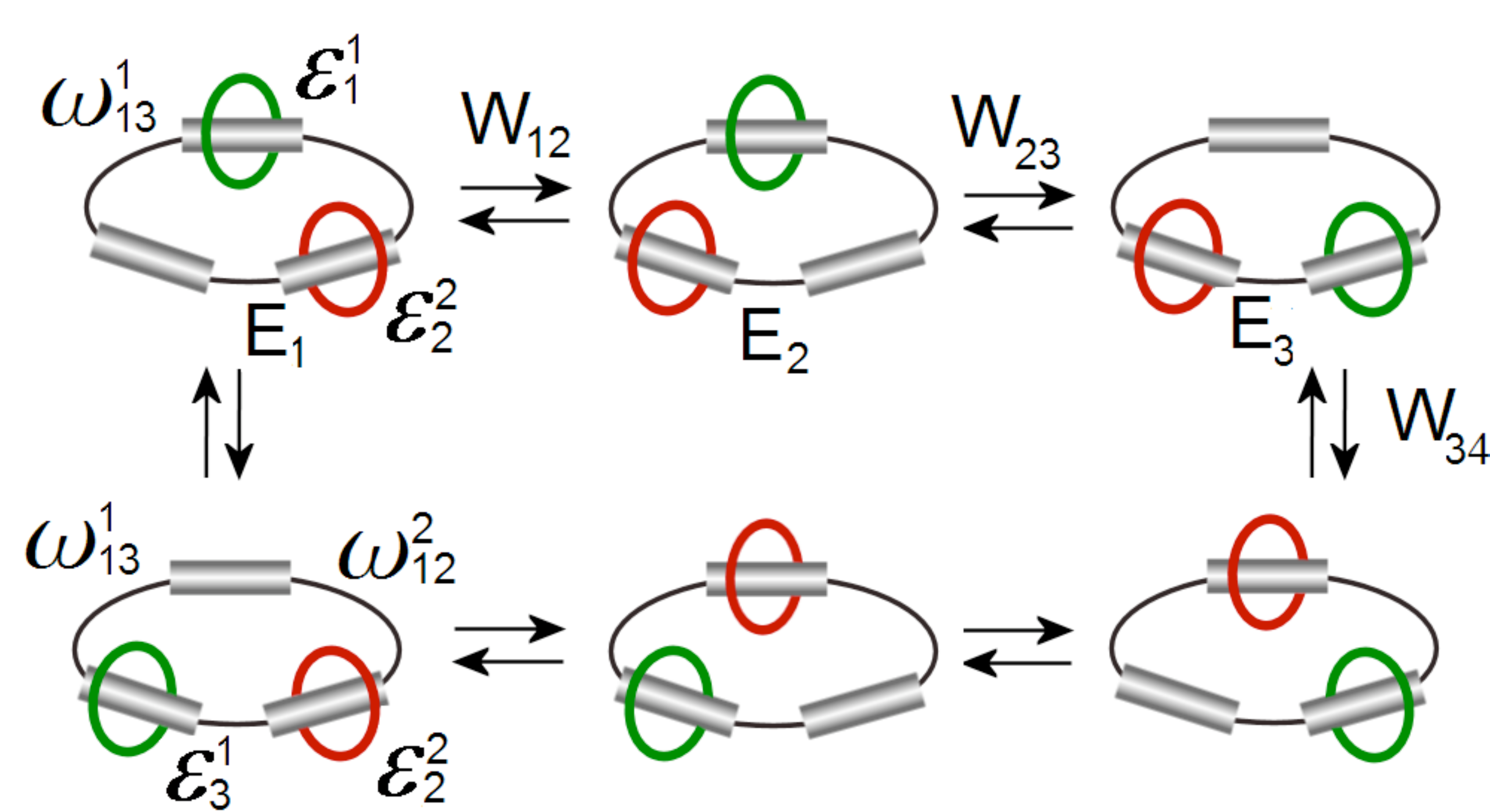}}
  \caption{Geometry and metastable states of a 3-catenane molecule.}
\label{3catenane}
\end{figure}

Molecular motors are molecules capable of performing controlled mechanical
motion.
%
An ability to rotate its parts is a crucial function of a molecular motor
\cite{astumian-07pnas,catenane,Leigh-03}. It is challenging to control this motion in strongly fluctuating
environment, experienced by any nanoscale system at room temperature. In the experiment \cite{Leigh-03}, controlled 
rotation was implemented using 2- and 3-catenane molecules that are made of interlocked polymer
rings (n-catenane is made of n  rings). Fig.~\ref{3catenane} shows  geometry of a 3-catenane molecule and its 6 metastable states.
Small (mobile) rings
perform  transitions among 3 stations on the third, larger, ring. These transitions
are caused by thermal fluctuations and, alone, may not lead to directed (clockwise or
counterclockwise) motion on average. In experiments, the directed motion was induced by modulating
the coupling strengths of mobile rings to stations. This forced the smaller rings to orbit around
the center of the larger ring while remaining interlocked with it.

In this Communication, we address an observation \cite{astumian-07pnas,Leigh-03,sinitsyn-09review,shi} that a  molecular motor can
perform robust quantized operations, e.g. making a full mobile ring rotation per cycle in the
control parameter space, even though the ring transitions are stochastic. 
We illustrate the phenomenon of integer quantization using a specific example of a 6-state
stochastic model in Fig.~\ref{3catenane}. The problem of finding the number of rotations can be
formulated in terms of stochastic motion on a graph $X$ whose nodes and edges represent the
metastable states and allowed transitions, respectively. The transition rates that satisfy the
detailed balance can be written in {\it Arrhenius form}, i.e. they can be parameterized by the well depths $E_i$ and potential barriers
$W_{ij}=W_{ji}$, so that the kinetic transition rate  from the node $j$ to $i$ is given by
$k_{ij}=e^{\beta(E_j-W_{ij})}$, with $\beta=1/k_BT$ being the inverse temperature. Even if the
rates can be written in Arrhenius form at any time, periodic changes of well depths and barriers result
in a directed particle motion. The phenomenon is referred to as the {\it stochastic pump effect}
\cite{sinitsyn-09review,sinitsyn-07epl,CS08,jarzynski-08prl}.

For the model in Fig.~\ref{3catenane}, we further express $W_{ij}$ and $E_i$ in terms of the
directly controllable parameters, represented by the coupling energies $\varepsilon_j^i$ of $i$-th mobile
ring to $j$-th station, or the potential barriers $\omega_{nm}^i$ between the station $n$, occupied
by the $i$-th ring, and an empty station $m$. By requesting
mobile rings be unable to occupy the same station simultaneously, and by comparing state energies and kinetic rates written in different parametrizations,
e.g. $E_1=\varepsilon_1^1+\varepsilon^2_2 $ or $e^{\beta(E_1-W_{16})}=e^{\beta(\varepsilon_1^1-\omega^1_{13})}$, we can relate the energies
$\varepsilon_j^i$ and barriers $\omega_{mn}^i$ to $E_k$ and $W_{kl}$ in an effective 6-state model
in Fig.~\ref{3catenane}, that is,
$E_2=\varepsilon_1^1+\varepsilon^2_3$, $E_3=\varepsilon_2^1+\varepsilon^2_3$, $E_4=\varepsilon^1_2+\varepsilon^2_1$,
$E_5=\varepsilon^1_3+\varepsilon^2_1$,
$E_6=\varepsilon^1_3+\varepsilon^2_2$, $W_{12}=\varepsilon^1_1+\omega^2_{23}$,
$W_{23}=\varepsilon^2_3+\omega^1_{12}$, $W_{34}=\varepsilon^1_2+\omega^2_{13}$, $W_{45}=\varepsilon^2_1+\omega^1_{23}$,
$W_{56}=\varepsilon^1_3+\omega^2_{12}$, and $W_{16}=\varepsilon^2_2+\omega^1_{13}$.

Consider a cyclic adiabatic evolution of $\varepsilon_1^1$ and $\varepsilon^1_2$.
 We introduce a {\it pump current vector} ${\bm Q}^{\bm s}$, whose components
 $Q_{ij}^{\bm s}$ are  average numbers of times the particle passed through  links  $(i,j)$ during the cycle ${\bm s}$ of control parameters evolution, i.e.
 ${\bm Q}^{\bm s}=\int_0^T{\bm J}(t) dt,$
where
$ {\bm J}(t)$ is the instantaneous current vector and $T$ is time of cyclic evolution of control parameters. The
 graph corresponding to Fig.~\ref{3catenane} has one loop and hence   $Q^{\bm s}_{i,i+1}=Q$ for any $i$, i.e. pump currents are the
same for each link.

\begin{figure}
\centerline{\includegraphics[width=2.8in]{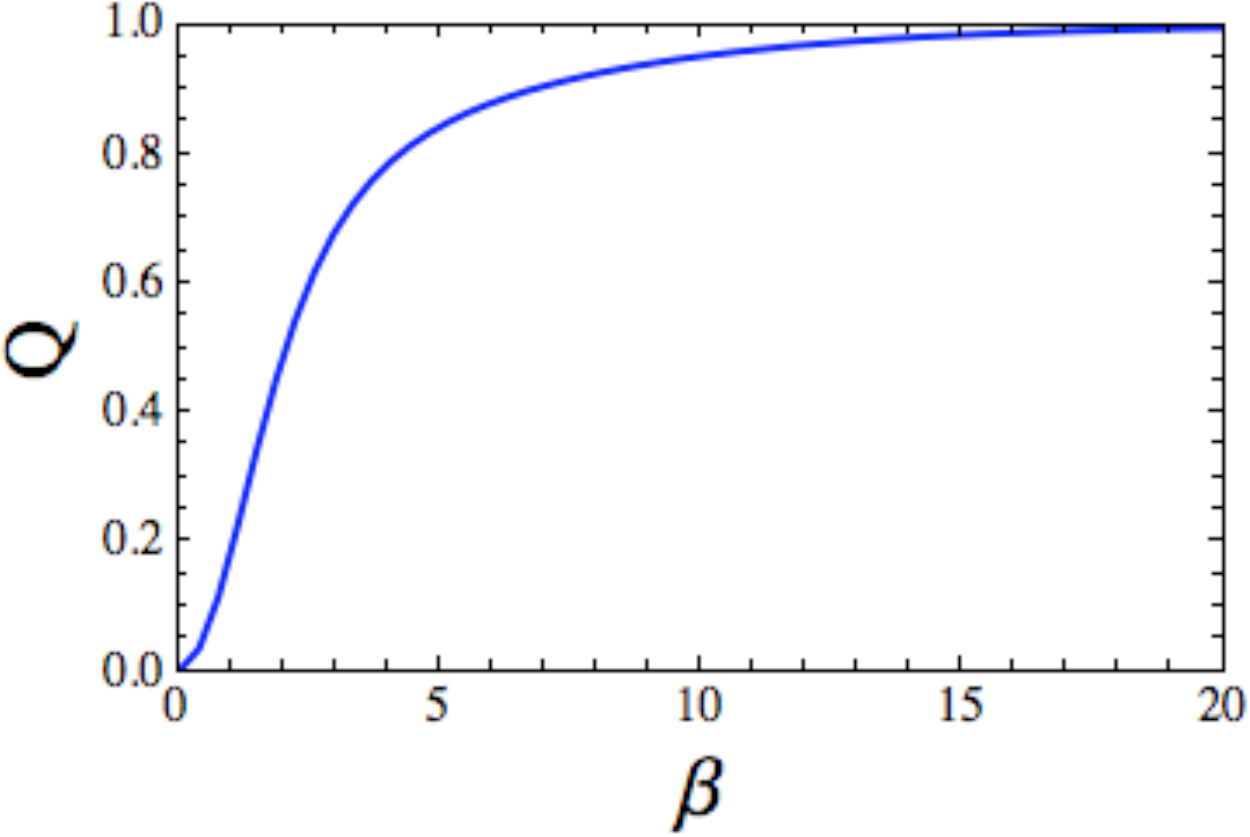}}
 \caption{Average number of rotations of mobile rings vs inverse temperature in a
3-catenane molecule after adiabatic evolution of control parameters along the contour
$\epsilon_1^1=-\epsilon_1^2=\cos(\phi)$, $\epsilon^1_2=-\epsilon^3_2=\sin(\phi)$, $\phi \in (0,2\pi)$;
 $\omega^2_{23}=1/4$, $\omega^{2}_{13}=1/2$, $\omega^2_{12}=-1/2$, $\omega^1_{12}=0$,
$\omega^1_{23}=-1/8$, $\omega_{13}^1=-1/7$, $\varepsilon^1_3=\varepsilon^2_3=0$.}
\label{quant}
\end{figure}

Fig.~\ref{quant}
shows the dependence of the number $Q$ of ring rotations per cycle on $\beta$, obtained by
solving the Master Equation in the adiabatic limit. It shows that, generally, the system's response
to a periodic parameter variation is not quantized, however, in the low-temperature $\beta
\rightarrow \infty$ limit, $Q$ saturates to an integer value $Q=1$. We have checked for a number of
models that the phenomenon is generic. By choosing arbitrary closed contour in the space of control
parameters, followed by choosing the remaining constant parameters randomly, integer response was
always achieved in the $\beta \rightarrow \infty$ limit.

{\it The Pumping-Quantization Theorem} rationalizes the observation and makes the following
assertion: Consider any finite
graph representing a Markov chain with kinetic rates written in Arrhenius form.  If during the cyclic  evolution of control parameters no degeneracy of the potential barriers
can encounter simultaneously with degeneracy of the minimal well depths, then in the adiabatic and
after this low temperature ($\beta\to\infty$) limits, the average number of particle transitions
through any link of a graph per a driving cycle is an integer. 

We emphasize that by considering $\beta\to\infty$ limit {\it after}
the adiabatic approximation we assume that the particle has sufficient time to explore the whole
phase space by making stochastic transitions before substantial change of control parameters can
happen. Hence the number of rotations of the system per cycle is random and the quantization, that we discuss, appears only on average.
Our arguments for the PQT are based on showing that various particle paths on a  graph, that
essentially contribute to the total current, are
{\it homologically equivalent}  to a single closed path, i.e. they differ from each other only by multiple transitions in {\it both}
 directions through some links. This is sufficient for a proof, since a single closed path can obviously pass through any link only an integer number of times.
 The proof of PQT includes three steps.


(i) 
 First, we prove the following identity 
\begin{equation}
\label{conditions-nabla} ({\bm J},{\bm J}^{\rm cons}) \equiv \sum_{\{i,j\}}e^{\beta W_{ij}}J_{ij}J^{\rm cons}_{ij}=0, 
\end{equation}
valid for the physical current ${\bm J}$ and any {\it conserved} current ${\bm J}^{\rm cons}$ that circulates in a loop of a graph and has equal values on any link of that loop.
Summation in (\ref{conditions-nabla}) runs over the graph links.
Eq.~(\ref{conditions-nabla}) follows from a fact that the physical
instantaneous  current can always be represented in the form
$J_{ij}=e^{-\beta W_{ij}}(e^{\beta E_j} \delta p_j-e^{\beta E_i} \delta p_i)$, where
$\delta p_i$ is deviation of the probability on the $i$-th node from its equilibrium value. Taking
any loop of a graph and summing $(e^{\beta E_j} \delta p_j-e^{\beta E_i} \delta p_i)$ along  its links gives zero, which is equivalent to (\ref{conditions-nabla}).

In the case of no barrier degeneracy,
due to the $e^{\beta W_{ij}}$
factor in (\ref{conditions-nabla}), the link $\{k,l\}$ with the largest barrier inside any loop on a graph dominates the scalar product
(\ref{conditions-nabla}) in the $\beta\to\infty$ limit, and the only way to make this product zero is to conclude that
$\lim_{\beta\to\infty}(J_{kl})=0$. Consider a time segment $\alpha$ without barrier degeneracies and let ${\bm Q}^{\alpha}$ be the mean current integrated over the time of this segment.
Integrating (\ref{conditions-nabla}) over time leads
to inequality $|Q_{kl}^{\alpha}| \le \sum_{\{i,j\} \ne \{k,l\}} {\rm max} \left( e^{\beta(W_{ij}-W_{kl})} \right) |Q_{ij}^{\alpha}|$, where  summation is over all links of the loop except the link $\{k,l\}$ with the highest barrier, and 
${\rm max} (...)$ means the maximum value of the expression during the given time segment.
The factors ${\rm max} \left( e^{\beta(W_{ij}-W_{kl})} \right)$ become 
infinitely small in the $\beta \rightarrow \infty$ limit but, for any given finite contour in the space of control parameters,
$|Q_{ij}^{\alpha}|$ remains finite even in the limit of infinitely 
long adiabatic evolution because pump currents, on average, depend only on a choice of a path of control parameters but do not depend explicitly on
time of the evolution \cite{sinitsyn-07epl,jarzynski-08prl}. This leads to a stronger  result $\lim_{\beta\to\infty} |Q_{kl}^{\alpha}| = 0$.

(ii) 
Here we note that the suppression of transitions through highest loop barriers during time interval without barrier degeneracies means that complex particle motion
 is restricted to a subgraph $\tilde{X}_{{\bm W}}\subset X$, referred to as the  {\it maximal spanning tree}, that depends only on the ordering
of non-degenerate barriers ${\bm W}$. It is constructed
from $X$ step-by-step by eliminating the edge with the largest barrier that does not destroy the
connectivity. Eventually, when none of the links can be removed without disconnecting the graph, we obtain
the tree $\tilde{X}_{\bm W}$. When energy $E_{j'}$ approaches the lowest energy $E_{j}$ 
and then becomes a new energy minimum, the particle travels from site $j$ to site $j'$. Let  $l_{jj'}(\tilde{X}_{\bm W})$ be the unique shortest path that connects $j$ to
$j'$ via $\tilde{X}_{{\bm W}}$. Although transitions form  $j$ to $j'$ are stochastic and can be done in a variety of ways,
all paths from $j$ to $j'$ on $\tilde{X}_{\bm W}$ are  homologically
equivalent to $l_{jj'}(\tilde{X}_{\bm W})$.


 (iii) To include events of barrier degeneracies,
 we partition the time period of a driving protocol into a set of small enough segments, so
that for each segment we either encounter only the minimal well depth degeneracy or just the
potential barrier degeneracy. We refer to them as to $0$- and $1$-segments, respectively. If
necessary, we merge the consecutive segments of the same type to make the $0$- and $1$-segments
 alternating.
In the $\beta \rightarrow \infty$ limit, no current is generated on the
$1$-segments, since the populations are concentrated  in the node
$j$ with the lowest value of $E_{j}$ but a nonzero on average pump current is possible only when the state probability vector changes \cite{jarzynski-08prl}. 
The populations at the beginning and at the end
of the 0-segment $\alpha$ are concentrated, respectively, in well-defined nodes $j_{\alpha}$ and $j'_{\alpha}$,
determined by the neighboring $1$-segments. Since the paths with nonvanishing probabilities from  $j_{\alpha}$ to $j'_{\alpha}$ belong to a tree $\tilde{X}_{{\bm W}_{\alpha}}$, the total current ${\bm Q}^{\alpha}$, passing during the time of the 0-segment $\alpha$ has values 1 on any link
 on the shortest path $l_{j_{\alpha}j'_{\alpha}}(\tilde{X}_{{\bm W}_{\alpha}})$
and it has zero values on other links. 
Finally, the current per cycle ${\bm Q}^{\bm s}=\sum_{\alpha}{\bm Q}^{\alpha}$ is generated by the concatenation of the
consecutive paths $l_{j_{\alpha}j'_{\alpha}}(\tilde{X}_{{\bm W}_{\alpha}})$ that correspond to the
$\alpha$-segments. This explicitly identifies the generated integer-valued current ${\bm Q}^{\bm s}$ and completes the proof of the PQT.

%
%
%
%

According to PQT, non-integer
quantization in 3-catenane molecules is highly unlikely when all mobile
rings and stations are different. In what
follows we argue that robust fractional quantization can occur in systems, where, due to some
symmetries, permanent (rigid) degeneracy of certain wells and/or barriers takes place.
%
%
\begin{figure}
\centerline{\includegraphics[width=1.8in]{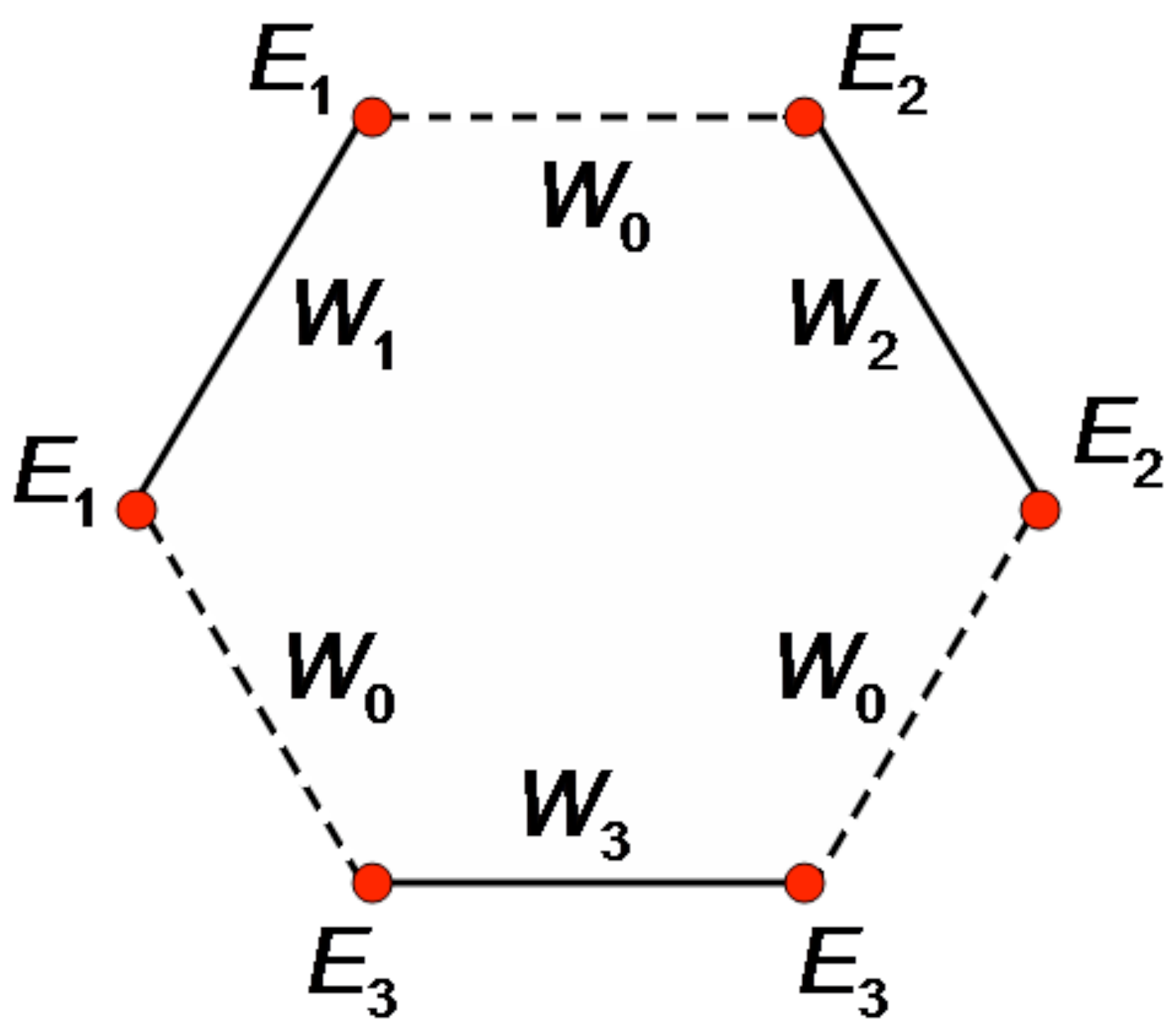}} \caption{The 6-state model with
triple degeneracy of barriers and double degeneracy of all well depths.}
\label{hexagon}
\end{figure}
%
To illustrate the point we consider the 3-catenane system in Fig.~\ref{3catenane}, with a special
symmetry: one of the mobile rings, called {\it active}, has three residence energies
$\varepsilon_{j}$, available for control, whereas the other, {\it passive}, mobile ring has the
same constant residing energy $\varepsilon=0$ for all three stations. All barriers for both rings
are constant and identical, equal to $\omega$. The system is described by the Markov-chain model in
Fig.~\ref{hexagon} defined on the cyclic $6$-node graph $X$ with permanently degenerate wells
$E_{2j-1}=E_{2j}=\varepsilon_{j}$, distinct barriers $W_{j}=\varepsilon_{j}+\omega$ with $j=1,2,3$,
and a permanently degenerate barrier $W_{0}=\omega$. The links $\{2j-1,2j\}$ and $\{2j,2j+1\}$
describe the transitions when the passive and the active ring, respectively, switches to the
non-occupied station.
%

The total current ${\bm Q}^{\bm s}$ can be calculated in a way similar to the integer-quantization
case by partitioning the time period into a set of alternating $0$- and $1$-  segments.
Consider a segment $\alpha$ with no degeneracy among the  parameters $W_{k}$ with $k=0,1,2,3$.
Similar to the integer-quantized case, the largest barriers still dominate the scalar product in Eq.~(\ref{conditions-nabla}), however, they can now be degenerate. Let $\Delta p_i^{\alpha}$ be the change
of the probability at site $i$ during the 0-segment $\alpha$ and let ${\rm max}_{\alpha}$ be the set of links $\{i,i+1\}$ with the largest barriers during $\alpha$. Then, in
the low-temperature limit, Eq.~(\ref{conditions-nabla}) combined with the continuity equations leads to
 \begin{equation}
\label{prod} \sum_{{\rm max}_{\alpha}} Q^{\alpha}_{i,i+1}=0, \quad \Delta p_j^{\alpha} = Q^{\alpha}_{j-1,j}- Q^{\alpha}_{j,j+1}, \quad j=1,\ldots, 5.
\end{equation}
Eqs. (\ref{prod}) completely determine currents ${\bm Q}^{\alpha}$ on each time segment.
Note that solution of Eqs. (\ref{prod}) results in generally rational values for $Q^{\alpha}_{j,j+1}$ because $\Delta p_j^{\alpha}$ take values in a set (-1/2,0,1/2) and all other
coefficients in equations are integers.

\begin{figure}
\centerline{\includegraphics[width=2.8 in]{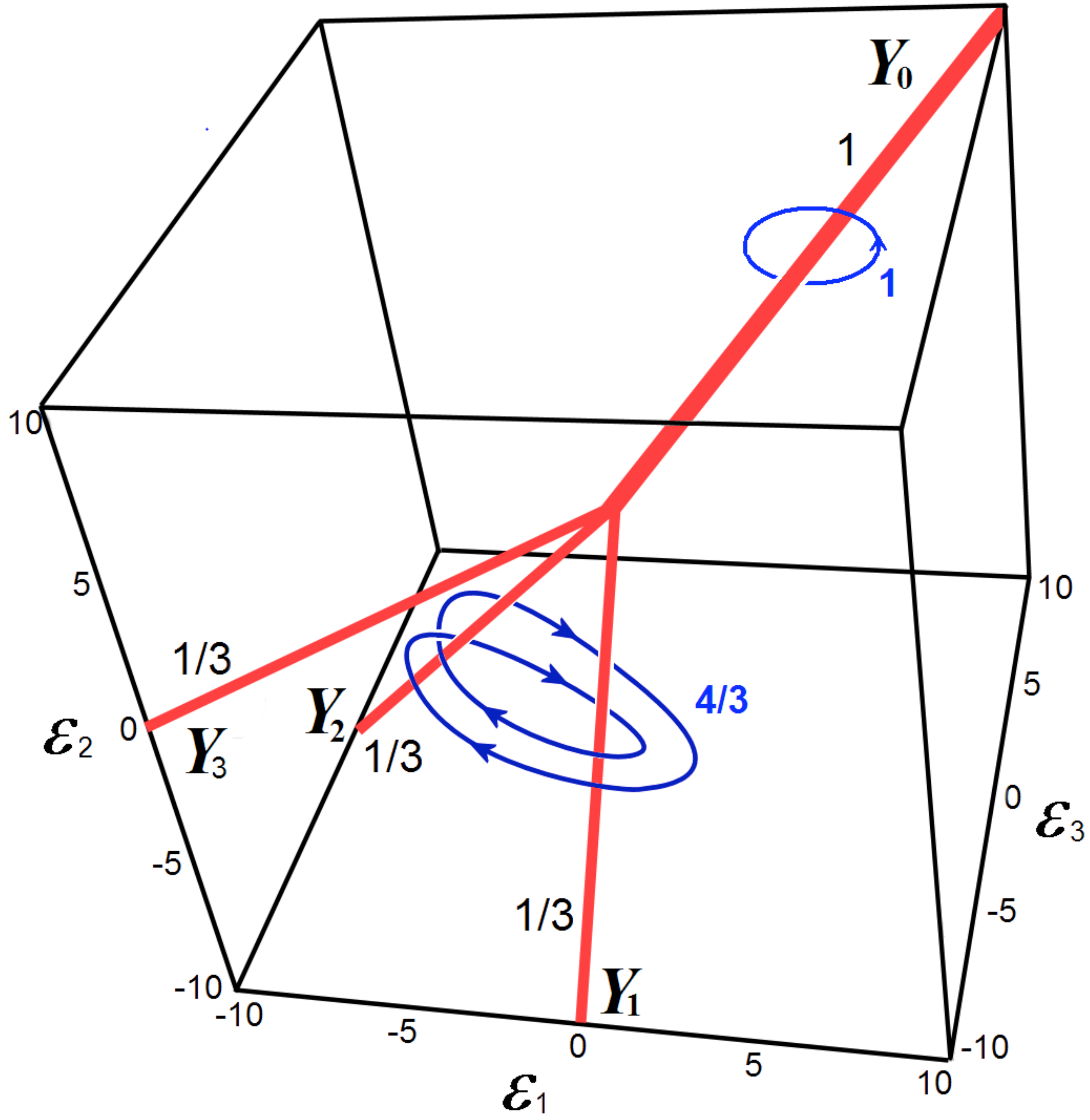}} \caption{Topology of the degeneracy space and currents in
 the active/passive ring model}
\label{tubes}
\end{figure}

The protocols under consideration avoid the set $Y$
of ``bad'' parameters, characterized by simultaneous degeneracy of the lowest wells and highest
barriers. For the  model in Fig.~\ref{hexagon},  $Y$  consists of $4$ lines $Y_{a}$, $a=0,1,2,3$, in the space of
${\bm\varepsilon}=(\varepsilon_{1},\varepsilon_{2},\varepsilon_{3})$, as it is shown in Fig.~\ref{tubes}. To understand the global
phase diagram of the quantized response, we examine the currents $Q_{a}^{s}$, the same for all links, that are generated by parameter motion along small contours
 enclosing $Y_{a}$ only one time. 
Finding the currents on all contributing 0-segments, as described above, and then summing them yields the currents of $1$ and $1/3$, as shown in Fig.~\ref{tubes}.
The independence of $Q_a$ on a variation of a contour around $Y_a$ allows the corresponding fluxes
${\cal Q}_{0}=1$ and ${\cal Q}_{j}=1/3$ for $j=1,2,3$ to be associated with the lines. It also suggests a topological nature of the
quantized current: the response to a general contour is given by the rational winding-index
$Q^{{\bm s}}=\sum_{a}n_{a}{\cal Q}_{a}$, where $n_{a}$ is the number of times the contour encloses the
line $Y_a$, taken with a proper sign depending on orientations (see Fig.~\ref{tubes}).

There is an obvious analogy between topological properties of pump currents and the Aharonov-Bohm effect in quantum mechanics. In the latter, the phase of the
electronic wavefunction  changes upon enclosing a quasi-1D solenoid with a magnetic field by an amount proportional to the total flux of the field inside the solenoid.
This analogy can be extended using the recent observation that stochastic pump effect is a geometric phenomenon \cite{sinitsyn-07epl,jarzynski-08prl}
 in a sense that integrated over time current can be written as a contour integral in the space of
control parameters $ Q^{{\bm s}} = \oint_{{\bm s}} {\bm A}\cdot d {\bm \varepsilon}$, where ${\bm A}$ is a vector potential (gauge field) in the space of control parameters.
This allows an effective ``magnetic field''  ${\bm B}={\bf \nabla}_{{\bm \varepsilon}}\times {\bm A}$ to be introduced. Our explicit calculations 
show that, at low temperatures, the field ${\bm B}$ is localized in narrow tubes, carrying fluxes $1$ and $1/3$, which become $Y_a$-lines in the $\beta \rightarrow \infty$ limit.

The arguments leading to rational quantization in our model with degeneracies
can be applied to any graph with rigid degeneracy of some barriers and/or potential wells.
For example, rigid degeneracies appear in the 3-catenane model with identical mobile rings.
Similar considerations  predict fractional quantization with a
minimal ratio $1/2$ for this system. We checked numerically that
this fractional quantization is robust when parameters are varied keeping mobile rings identical.
However, it is destroyed as soon as mobile rings are made different.

In conclusion, we have shown that current quantization in a stochastic system is a generic
phenomenon and we identified the conditions for its observation. PQT directly applies to experiments
with catenane molecules predicting integer-quantized response in a generic situation. We showed,
however, that additional symmetries can lead to fractional quantization. For example, we predict
1/2 quantization in a 3-catenane molecule that has two identical rings. We also showed that a
3-catenane molecule can demonstrate fractional quantization with a minimal ratio of 1/3, which has
not been observed previously. Quantization of a molecular motor response is topologically
protected. This robustness should have applications to the control of nanoscale systems,
experiencing thermal fluctuations. 



\begin{acknowledgments}
{\it We are grateful to M. Chertkov, J. R. Klein and J. Horowitz for useful discussions. This material is based upon work supported by NSF under Grants No. CHE-0808910 and ECCS-0925618.}
\end{acknowledgments}

\newpage

\appendix

\end{document}